\documentclass[a4paper,10pt,twoside]{cpc-hepnp}

\usepackage{supertabular,booktabs}
\usepackage{color}
\usepackage{multirow}
\usepackage{comment}
\usepackage[pagewise]{lineno}
\usepackage{longtable}
\usepackage{multicol}
\usepackage{graphicx}
\usepackage{booktabs}
\usepackage{amssymb,bm,mathrsfs,bbm,amscd}
\usepackage[tbtags]{amsmath}
\usepackage{lastpage}
\usepackage{CJK}
\begin{document}
\begin{CJK*}{GBK}{song}


\title{Measurement of the Integrated Luminosities of Cross-section Scan Data Samples Around the $\psi(3770)$ Mass Region}
\maketitle
\begin{small}
\begin{center}
M.~Ablikim$^{1}$, M.~N.~Achasov$^{9,d}$, S. ~Ahmed$^{14}$, M.~Albrecht$^{4}$, M.~Alekseev$^{55A,55C}$, A.~Amoroso$^{55A,55C}$, F.~F.~An$^{1}$, Q.~An$^{52,42}$, Y.~Bai$^{41}$, O.~Bakina$^{26}$
% ...
\\
\vspace{0.2cm}
(BESIII Collaboration)\\
\vspace{0.2cm} {\it
$^{1}$ Institute of High Energy Physics, Beijing 100049, People's Republic of China\\
% ...
$^{4}$ Bochum Ruhr-University, D-44780 Bochum, Germany\\
% ...
$^{9}$ G.I. Budker Institute of Nuclear Physics SB RAS (BINP), Novosibirsk 630090, Russia\\
% ...
$^{14}$ Helmholtz Institute Mainz, Johann-Joachim-Becher-Weg 45, D-55099 Mainz, Germany\\
% ...
$^{26}$ Joint Institute for Nuclear Research, 141980 Dubna, Moscow region, Russia\\
% ...
$^{41}$ Southeast University, Nanjing 211100, People's Republic of China\\
$^{42}$ State Key Laboratory of Particle Detection and Electronics, Beijing 100049, Hefei 230026, People's Republic of China\\
% ...
$^{52}$ University of Science and Technology of China, Hefei 230026, People's Republic of China\\
% ...
$^{55}$ (A)University of Turin, I-10125, Turin, Italy; (B)University of Eastern Piedmont, I-15121, Alessandria, Italy; (C)INFN, I-10125, Turin, Italy\\
% ...
$^{d}$ Also at the Novosibirsk State University, Novosibirsk, 630090, Russia\\
% ...
}\end{center}

\vspace{0.4cm}
\end{small}


\begin{abstract}
To investigate the nature of the $\psi(3770)$ resonance and to measure the cross section for $e^+e^- \to D\bar{D}$,
a cross-section scan data sample, distributed among 41 center-of-mass energy points from 3.73 to 3.89~GeV, was taken with the BESIII detector operated at the BEPCII collider in the year 2010.
By analyzing the large angle Bhabha scattering events, we measure the integrated
luminosity of the data sample at each center-of-mass energy point.
 The total integrated luminosity of the data sample is $76.16\pm0.04\pm0.61$~pb$^{-1}$, where the first uncertainty is
statistical and the second systematic.
\end{abstract}

\begin{multicols}{2}
\section{Introduction}
The $\psi(3770)$ is the lowest
mass charmonium state above the $D\bar{D}$ threshold, and
is generally regarded as the $1^{3}D_{1}$ dominant charmonium
state~\cite{3770}.
To investigate the nature of the $\psi(3770)$ resonance, the BESIII Collaboration
performed a
cross-section scan experiment, in which $e^+e^-$ data at 41 center-of-mass (CM) energy ($E_{\rm cm}$) points from
3.73 to 3.89~GeV were collected.
This data sample, referred to as the ``$\psi(3770)$ cross-section scan data,'' was collected during the time period from June 1st to June 16th, 2010.

The $\psi(3770)$ cross-section scan data can be used to study the line-shapes of the cross sections for
various hadronic final states produced in $e^+e^-$ annihilation in the energy region around the $\psi(3770)$.
Amplitude analyses of these line-shapes of cross sections will provide crucial information
to explore the anomalous line-shape observed by the BESII experiment in 2008~\cite{Int_ref2}.
 These also benefit the measurements of the parameters of the $\psi(3770)$ resonance and shed light on the understanding of the branching fraction of $\psi(3770)\to$ non-$D\bar D$~\cite{Int_nonDD1,Int_nonDD2,Int_nonDD3,Int_nonDD4,nonDDCLEO} decays.

In this paper, we present measurements of the integrated luminosity of the $\psi(3770)$ cross-section scan data at each $E_{\rm cm}$
by analyzing large angle Bhabha scattering events.
We follow a method similar to that used in the measurement of the integrated luminosity of the data taken at $E_{\rm cm}=$ 3.773 GeV with the BESIII detector~\cite{psi3770_lum}. Furthermore, the luminosities are checked with an independent measurement by
analyzing $e^+e^-\rightarrow(\gamma)\gamma\gamma$~events.
\section{BESIII detector}
BEPCII~\cite{ref1} is a double-ring $e^{+}e^{-}$ collider. The design peak luminosity is $1\times10^{33}$ cm$^{-2}$s$^{-1}$ at a
beam current of $0.93$ A and was achieved in 2016. The BESIII detector~\cite{ref1} has a geometrical acceptance of $93\%$ of $4\pi$ and consists of the following main
components: 1) a small-celled, helium-based main drift chamber (MDC) with 43 layers. The average single wire resolution
is 135 $\mu$m, and the momentum resolution for $1~\rm{GeV}$$/c$ charged particles in a $1~\rm{T}$ magnetic field is $0.5\%$; 2) an electromagnetic
calorimeter (EMC) made of 6240 CsI (Tl) crystals arranged in a cylindrical shape (barrel) plus two endcaps. For 1.0 GeV photons,
the energy resolution is $2.5\%$ (5\%) in the barrel (endcaps), and the position resolution is 6 mm (9 mm) in the barrel (endcaps); 3) a Time-Of-Flight system (TOF) for particle identification composed of a barrel part made of two layers with
88 pieces of 5 cm thick, 2.4 m long plastic scintillators in each layer, and two endcaps with 96 fan-shaped, 5 cm thick, plastic
scintillators in each endcap. The time resolution is 80 ps (110 ps) in the barrel (endcaps), corresponding to a $2\sigma$ K/$\pi$
separation for momentum up to about 1.0 GeV/$c$; 4) a muon chamber system (MUC) made of 1600 m$^{2}$ of Resistive Plate Chambers (RPC) arranged
in 9 layers in the barrel and 8 layers in the endcaps and incorporated in the return iron of the superconducting magnet. The position resolution is about 2 cm.
\section{Method}
In principle, any process with a well-known cross-section can be used to determine the integrated luminosity of the corresponding data set. The luminosity $\mathcal L$ can be calculated by
\begin{linenomath*}
\begin{equation}\label{eq:lum}
\mathcal L=\frac{N^{\rm obs}\times(1-\eta)}{\sigma\times\varepsilon},
\end{equation}
\end{linenomath*}
where $N^{\rm obs}$ is the number of observed events, $\eta$ is the background contamination rate, $\sigma$ is the cross section and $\varepsilon$ is the detection efficiency.

In $e^+e^-$ experiments, useful processes for
the determination of integrated luminosity are the QED processes $e^+e^-\to(\gamma)e^+e^-$,
$e^+e^-\to(\gamma)\gamma\gamma$ and $e^+e^-\to(\gamma)\mu^+\mu^-$
since they have precisely calculated cross sections in QED and relatively simple and distinctive final states. According to its largest production cross section,
the Bhabha scattering process~($e^+e^-\to(\gamma)e^+e^-$) is used to measure the integrated luminosity of
the $\psi(3770)$ cross-section scan data.
In this work, Babayaga v3.5~\cite{babayaga} is adopted as the generator to
determine the cross sections and the detection efficiencies.
\section{Luminosity measurement}
\subsection{Event selection}\label{sec:lummea}
The Bhabha scattering candidate events are selected by requiring exactly two oppositely-charged tracks that are well reconstructed in the MDC and satisfy $|\cos\theta|<0.70$,
where $\theta$ is the polar angle of the charged track.
Each good charged track must satisfy $|V_r|<1$~cm and $|V_z|<5$~cm.
Here $V_r$ and $V_z$ are the closest distance of the charged tracks to the interaction point in the plane perpendicular to the beam direction and along the beam direction, respectively.

To suppress the backgrounds from
$e^+e^-\to J/\psi\rm X$, where the $J/\psi$ decays into a $e^+e^-$ pair, and X refers to $\gamma_{\rm{ISR}}$, $\pi^0\pi^0$, $\eta$, $\pi^0$, or $\gamma\gamma$, the sum of the momenta of the two good charged tracks is required to be greater
than $0.9\times E_{\rm cm}/c$.
The momentum of each good charged track is also required to be less than $(E_{\rm{b}}/c+0.15)~\rm{GeV}$$/c$, where $E_{\rm b}$ is the beam energy and 0.15
GeV/$c$ is about 4 times the momentum resolution~\cite{psi3770_lum}.
The energy deposited in the EMC of each charged track ($E_{\rm EMC}$) is required to be larger than 1 GeV to reject the background from $e^+e^-\rightarrow(\gamma)\mu^+\mu^-$.

After applying the above selection criteria, most of the surviving events come from the process  $e^+e^-\to(\gamma)e^+e^-$.
Taking $E_{\rm cm}=3.7358$~GeV as an example,
comparisons of the distributions of the momentum, polar angle and deposited energy in the EMC of the charged tracks between data and Monte Carlo (MC) simulation are shown in Fig.~\ref{fig:cmp}.  Good agreement between data and MC simulation is observed.
\subsection{Background estimation}
Most of the surviving candidate events are from $e^+e^-\to(\gamma)e^+e^-$. Potential background contamination includes two parts. One is the beam-associated background, such as beam-gas and beam-wall events. The other is background from $e^+e^-$ reaction including $\psi(3770)\to D\bar{D}$, $\psi(3770)\to$ non-$D\bar{D}$,
$e^+e^-\to (\gamma)J/\psi$, $(\gamma)\psi(3686)$, $q\bar{q}$, $(\gamma)\mu^+\mu^-$ and $(\gamma)\tau^+\tau^-$. To study the beam-associated backgrounds, we analyzed the separated-beam data samples collected at 3.400 GeV and 4.030 GeV with BESIII.
To estimate the background contamination rates for the other background processes, we analyze large MC samples generated at $E_{\rm cm}=3.773$~GeV.
The overall contamination rate $\eta$ is estimated by
\begin{linenomath*}
\begin{equation}
 \eta=\frac{\sum\sigma^i\times\eta^i}{\sigma^{\rm Bhabha}\times\varepsilon^{\rm Bhabha}},
\end{equation}
\end{linenomath*}
where $\sigma^i$ and $\eta^i$ are the cross section and the contamination rate for a specific process $i$, respectively; and $\sigma^{\rm Bhabha}$ and $\varepsilon^{\rm Bhabha}$ are the cross section and detection efficiency, respectively, for the Bhabha scattering process.
The overall contamination rate of these backgrounds is estimated to be at the level of $10^{-4}$.
\subsection{Numerical result}
\label{sec:rslt}
Inserting the numbers of observed Bhabha scattering events,
the contamination rates of backgrounds, the detection efficiencies
 and cross sections calculated with the Babayaga v3.5 generator~\cite{babayaga} into Eq.~(\ref{eq:lum}),
we obtain the integrated luminosity at individual CM energy points for the $\psi(3770)$ cross-section scan data.

The measured integrated luminosities are summarized in the second column of Table~\ref{tab:rslt}. The total integrated luminosity of the $\psi(3770)$ cross-section scan data is determined to be $76.16\pm0.04\pm0.61$~pb$^{-1}$, where the first uncertainty is statistical and the second systematic, which will be discussed in the following.
\end{multicols}
\begin{center}
  \includegraphics[width=0.4\textwidth]{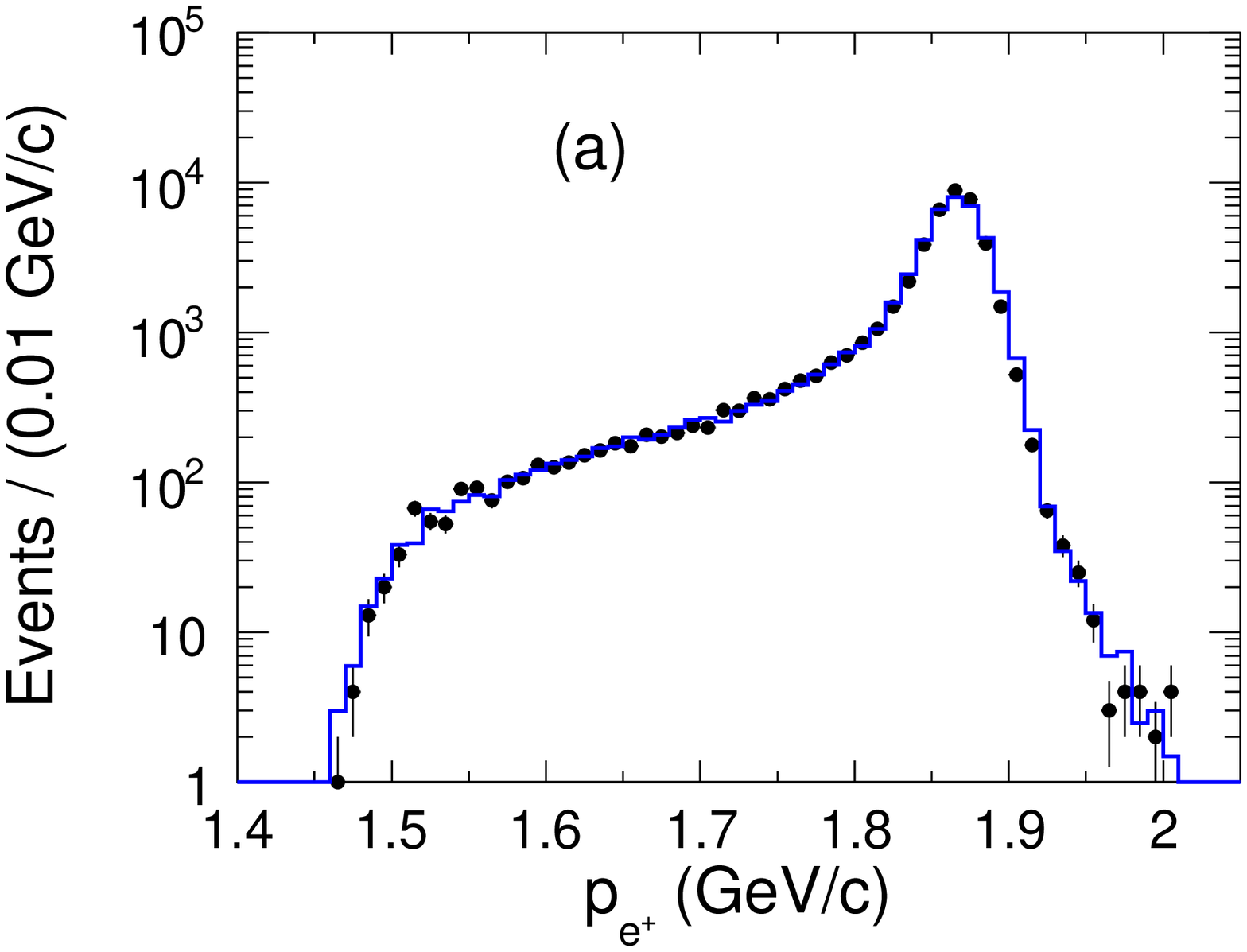}
  \includegraphics[width=0.4\textwidth]{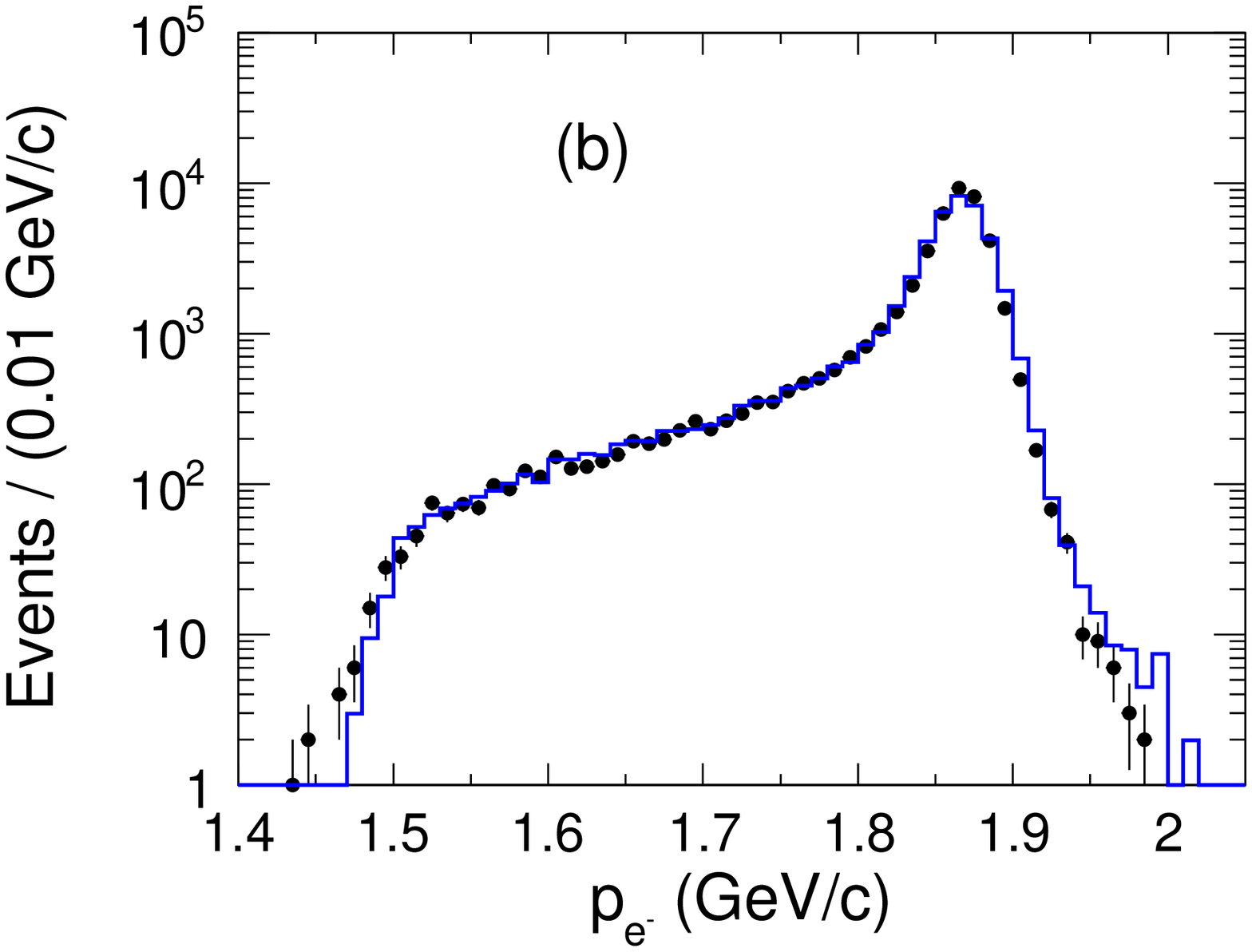}\\
  \includegraphics[width=0.4\textwidth]{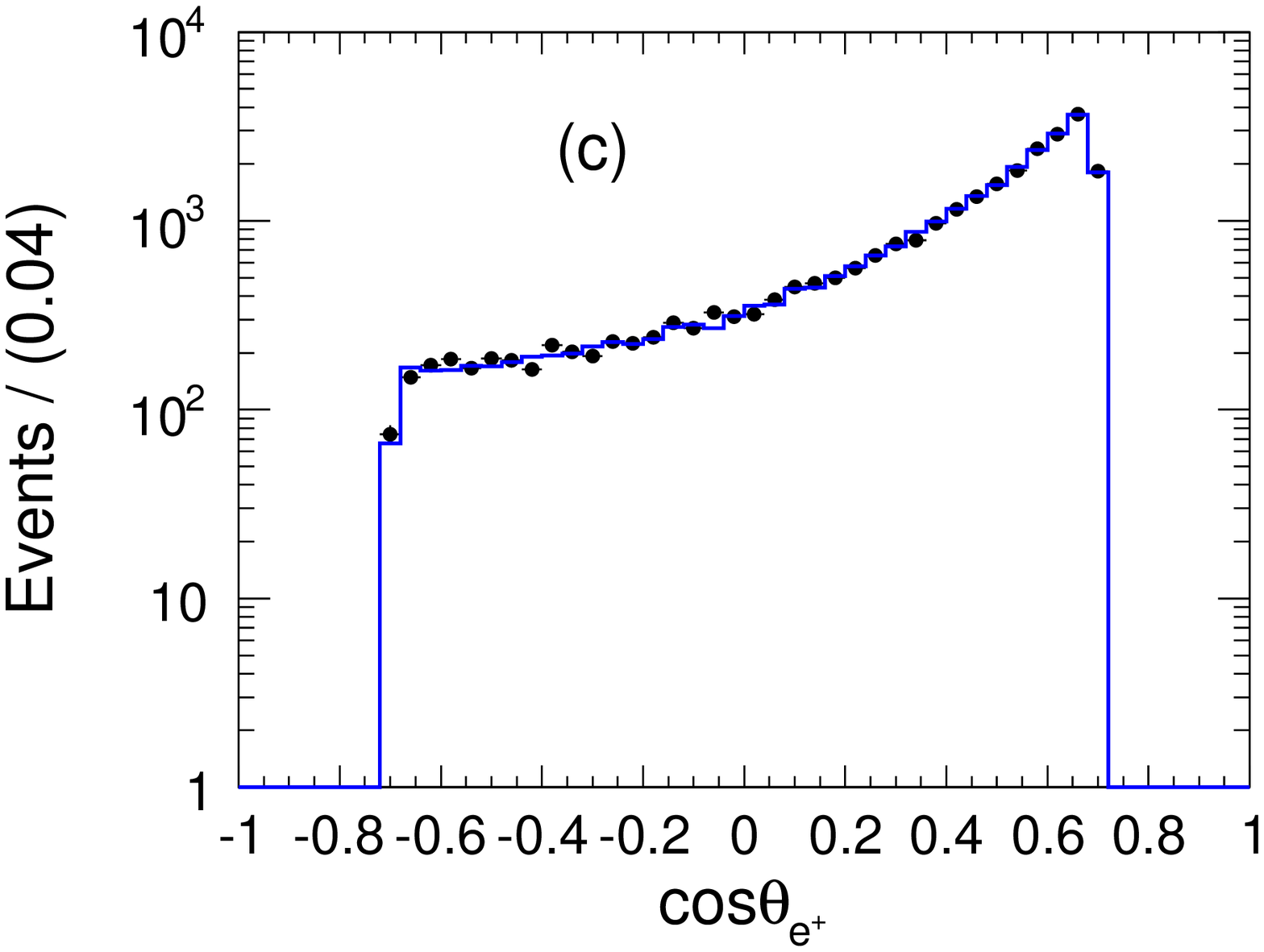}
  \includegraphics[width=0.4\textwidth]{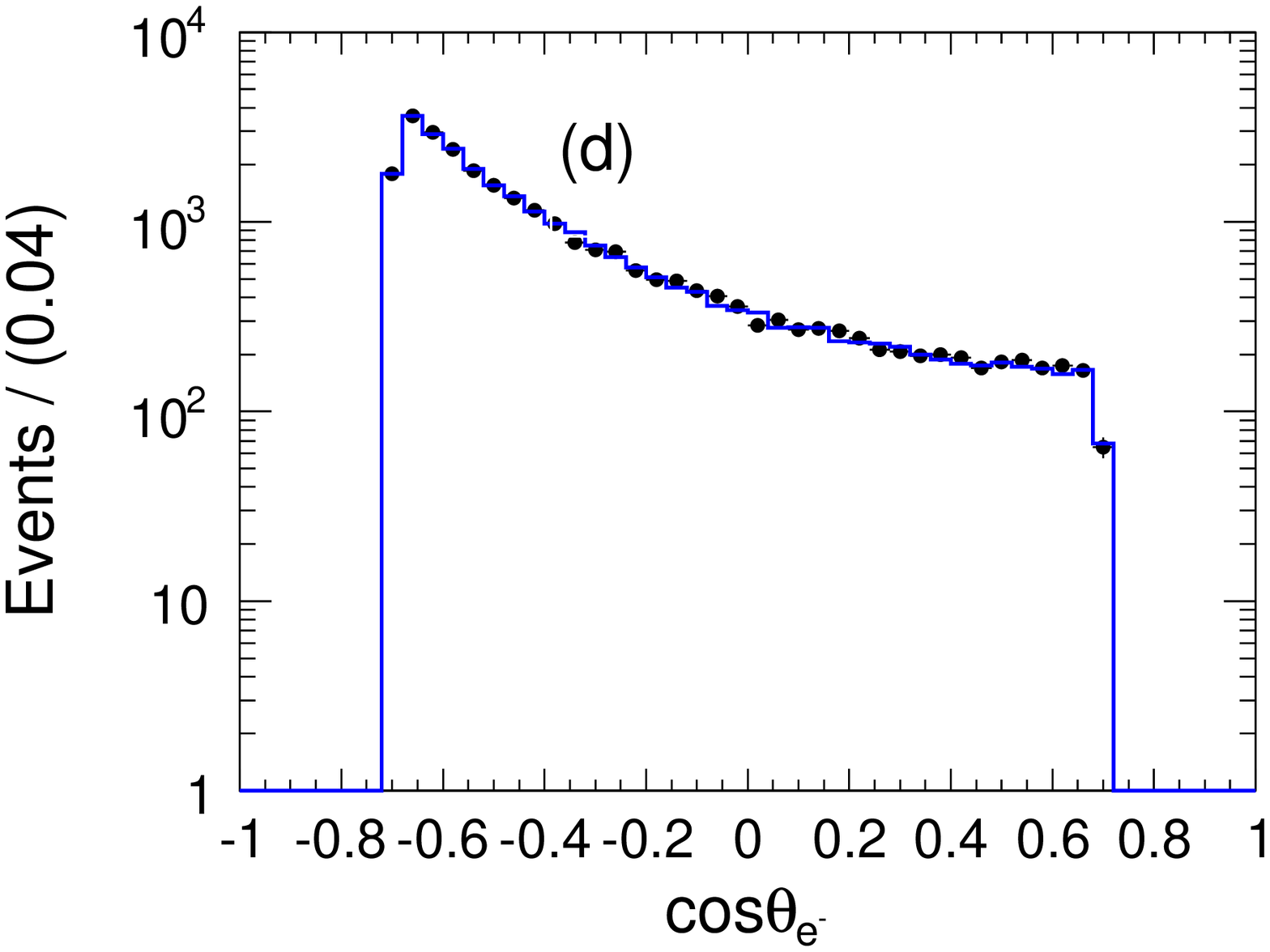}\\
  \includegraphics[width=0.4\textwidth]{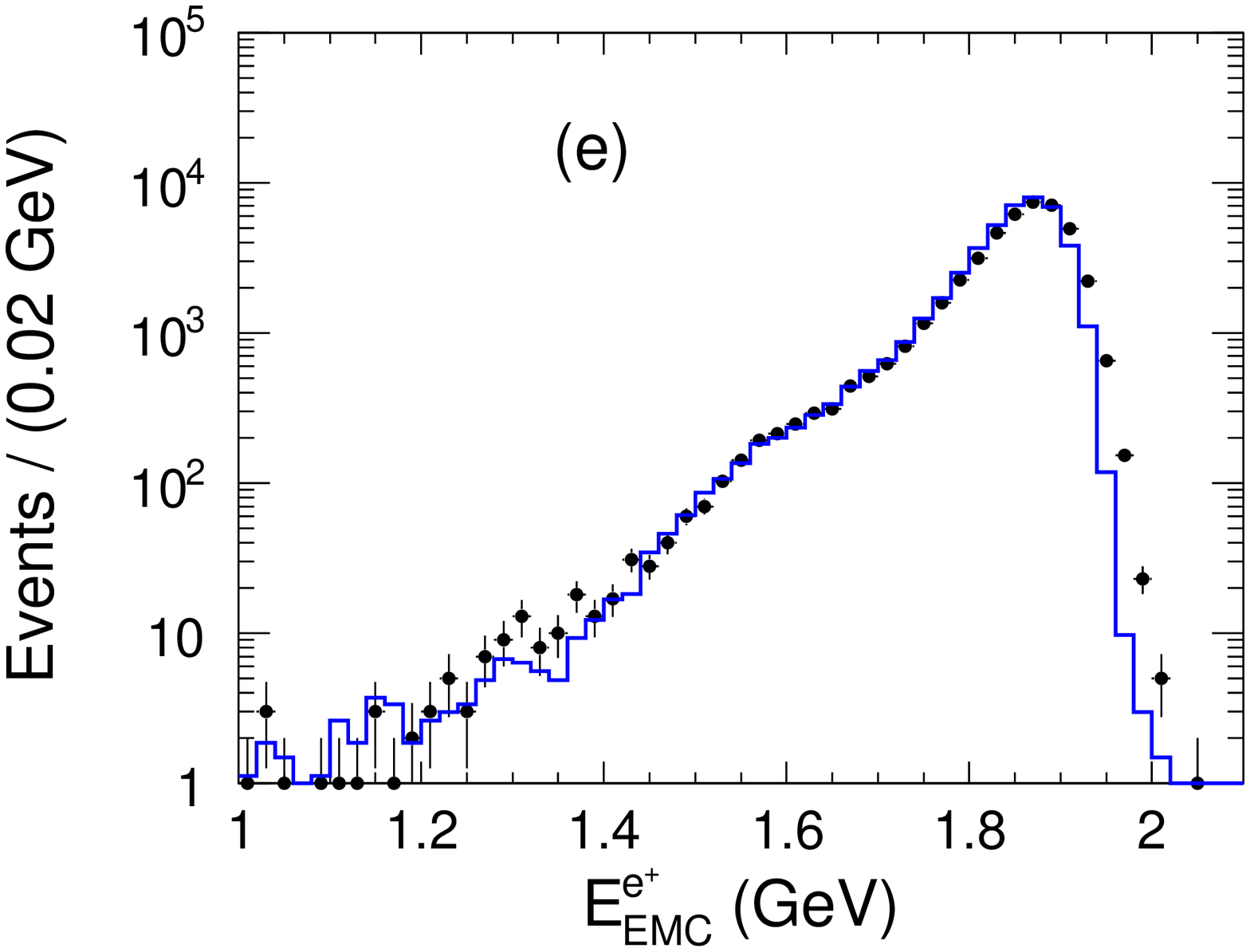}
  \includegraphics[width=0.4\textwidth]{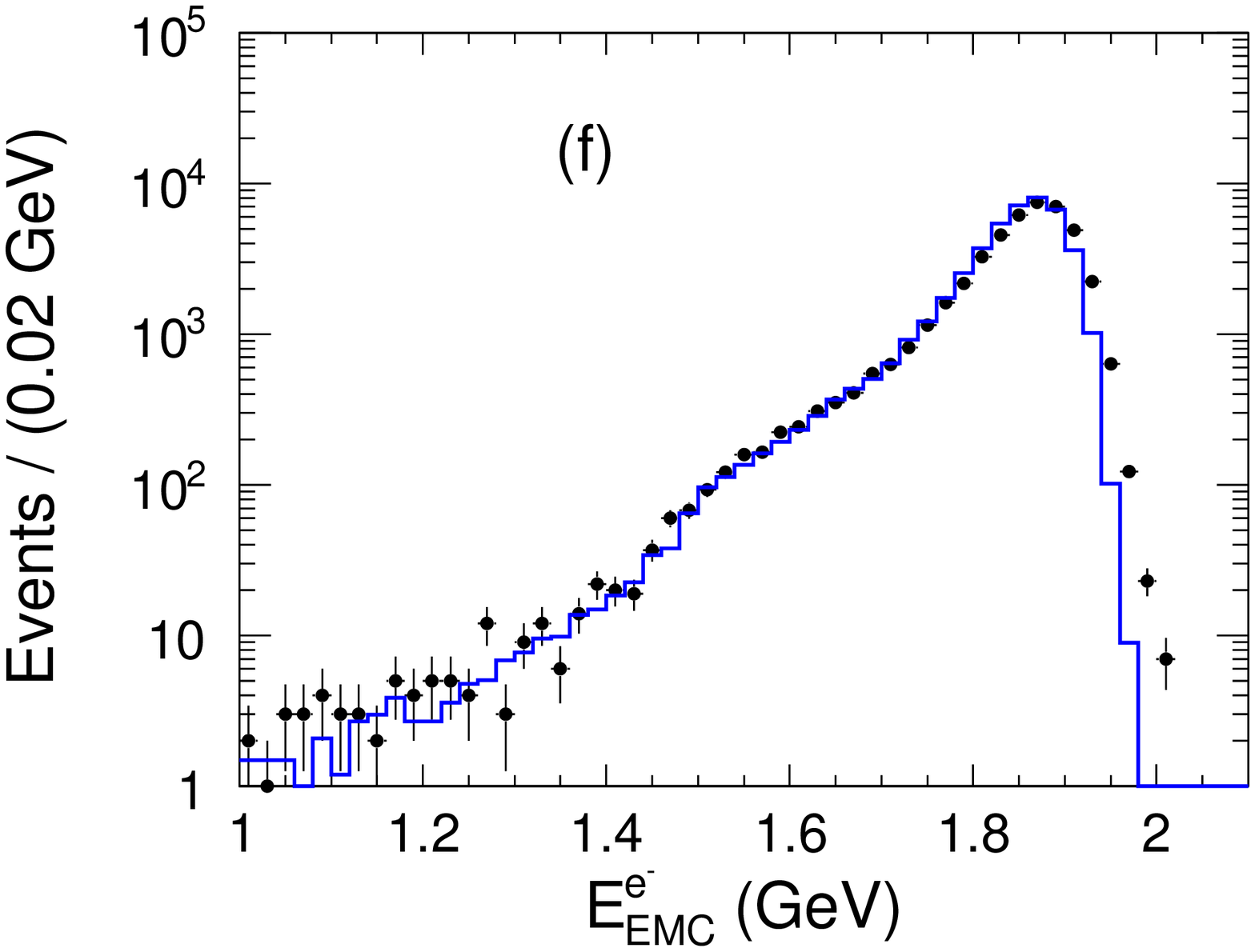}
\figcaption{\label{fig:cmp} Distributions of (a), (b) momentum, (c), (d) $\cos\theta$ and (e), (f) deposited energy in the EMC of the two charged tracks in the CM frame for selected Bhabha candidate events from the data taken at $E_{\rm cm}$ =3.7358 GeV (points with error bars) and the corresponding MC simulation (histograms). The MC entries are normalized to the experimental data.}
\end{center}
\begin{multicols}{2}
\section{Systematic uncertainty}
\label{sec:syserr}
The main sources of the systematic uncertainty are the event selection, the trigger efficiency, the generator, and the beam energy.
Due to the low luminosity of individual data sets, we take the average value among the 41 CM energy points as the systematic uncertainties to avoid large statistical fluctuations.

To estimate the systematic uncertainty of the $\cos\theta$ requirement, we repeat the measurements with the alternative requirements $|\cos\theta|<0.60$, $|\cos\theta|<0.65$, $|\cos\theta|<0.75$, or $|\cos\theta|<0.80$, individually. The maximum relative change of the total integrated luminosity with respect to the nominal value is taken as the systematic uncertainty.

To study the systematic uncertainty arising from the MDC information, including the tracking and momentum requirements, we select a Bhabha sample using only EMC information. Two clusters must be reconstructed in the EMC with a deposited energy larger than $0.85\times E_{\rm{b}}$ and a polar angle within $|\cos\theta|<0.7$. To remove $e^+e^-\rightarrow(\gamma)\gamma\gamma$ events, an additional requirement of $5^{\circ}<|\Delta\phi|<22^{\circ}$ is imposed, where $\Delta\phi$ is defined as $\Delta\phi=|\phi_{1}-\phi_{2}|-180^{\circ}$, and $\phi_{1}$ and $\phi_{2}$ are the azimuthal angles of the two showers in the EMC. The requirements on the MDC information are then imposed on the selected candidates, and the ratio of the surviving events is regarded as the corresponding acceptance efficiency. The difference of the acceptance efficiencies between data and MC simulation is taken as the relevant systematic uncertainty.
\begin{center}
\tabcaption{ \label{tab:rslt}  Summary of integrated luminosities measured using the processes  $e^+e^-\rightarrow(\gamma)e^+e^-$ ($\mathcal L^{e^+e^-}$) and $e^+e^-\rightarrow(\gamma)\gamma\gamma$ ($\mathcal L^{\gamma\gamma}$) at each individual CM energy, where the first uncertainties are statistical and the second are systematic.}
\footnotesize
\begin{tabular*}{80mm}{c@{\extracolsep{\fill}}cc}
\toprule $E_{\rm cm}$ (GeV)   & $\mathcal L^{e^+e^-}$ (nb$^{-1}$) & $\mathcal L^{\gamma\gamma}$ (nb$^{-1}$)\\
\hline
 3.6471  &$  2255.4\pm     6.3\pm    18.0$ &$   2250.3\pm15.5\pm24.8  $\\
 3.6531  &$  2214.0\pm     6.3\pm    17.7$ &$   2184.1\pm15.3\pm24.0  $\\
 3.7266  &$   896.2\pm     4.1\pm     7.2$ &$   879.8\pm9.9\pm9.7     $\\
 3.7356  &$   334.8\pm     2.5\pm     2.7$ &$   340.9\pm6.2\pm3.7     $\\
 3.7358  &$   491.9\pm     3.0\pm     3.9$ &$   484.8\pm7.4\pm5.3     $\\
 3.7376  &$   327.7\pm     2.5\pm     2.6$ &$   324.1\pm6.0\pm3.6     $\\
 3.7447  &$   956.0\pm     4.2\pm     7.6$ &$   933.9\pm10.3\pm10.3   $\\
 3.7464  &$  1412.2\pm     5.1\pm    11.3$ &$   1404.1\pm12.6\pm15.4  $\\
 3.7488  &$  2270.9\pm     6.5\pm    18.2$ &$   2267.6\pm16.0\pm24.9  $\\
 3.7503  &$  2971.8\pm     7.5\pm    23.8$ &$   2962.7\pm18.3\pm32.6  $\\
 3.7526  &$  3310.7\pm     7.9\pm    26.5$ &$   3308.1\pm19.4\pm36.4  $\\
 3.7541  &$  3418.1\pm     8.0\pm    27.3$ &$   3370.0\pm19.6\pm37.1  $\\
 3.7555  &$  3878.0\pm     8.5\pm    31.0$ &$   3824.9\pm20.9\pm42.1  $\\
 3.7585  &$  4444.8\pm     9.2\pm    35.6$ &$   4411.9\pm22.4\pm48.5  $\\
 3.7616  &$  4494.7\pm     9.2\pm    36.0$ &$   4456.9\pm22.5\pm49.0  $\\
 3.7645  &$  3290.3\pm     7.9\pm    26.3$ &$   3277.4\pm19.3\pm36.1  $\\
 3.7675  &$  2449.9\pm     6.8\pm    19.6$ &$   2419.2\pm16.6\pm26.6  $\\
 3.7705  &$  2021.7\pm     6.2\pm    16.2$ &$   2001.7\pm15.1\pm22.0  $\\
 3.7735  &$  1833.0\pm     5.9\pm    14.7$ &$   1818.0\pm14.4\pm20.0  $\\
 3.7765  &$  1829.4\pm     5.9\pm    14.6$ &$   1823.1\pm14.5\pm20.1  $\\
 3.7795  &$  1956.1\pm     6.1\pm    15.6$ &$   1933.1\pm14.9\pm21.3  $\\
 3.7825  &$  2148.3\pm     6.4\pm    17.2$ &$   2116.8\pm15.6\pm23.3  $\\
 3.7855  &$  2546.7\pm     7.0\pm    20.4$ &$   2538.0\pm17.1\pm27.9  $\\
 3.7882  &$  2840.9\pm     7.4\pm    22.7$ &$   2811.2\pm18.0\pm30.9  $\\
 3.7925  &$  3537.2\pm     8.2\pm    28.3$ &$   3506.3\pm20.1\pm38.6  $\\
 3.7964  &$  4056.9\pm     8.8\pm    32.5$ &$   4006.1\pm21.6\pm44.1  $\\
 3.8002  &$  3931.2\pm     8.7\pm    31.4$ &$   3911.1\pm21.3\pm43.0  $\\
 3.8026  &$  2690.5\pm     7.2\pm    21.5$ &$   2671.3\pm17.6\pm29.4  $\\
 3.8064  &$  1762.4\pm     5.8\pm    14.1$ &$   1732.0\pm14.2\pm19.1  $\\
 3.8095  &$  1252.3\pm     4.9\pm    10.0$ &$   1275.1\pm12.2\pm14.0  $\\
 3.8124  &$   898.5\pm     4.2\pm     7.2$ &$   898.5\pm10.3\pm9.9    $\\
 3.8156  &$   683.0\pm     3.6\pm     5.5$ &$   666.6\pm8.8\pm7.3     $\\
 3.8236  &$   399.5\pm     2.8\pm     3.2$ &$   386.3\pm6.7\pm4.2     $\\
 3.8315  &$   281.7\pm     2.3\pm     2.3$ &$   278.5\pm5.7\pm3.1     $\\
 3.8396  &$   282.3\pm     2.4\pm     2.3$ &$   269.6\pm5.7\pm3.0     $\\
 3.8475  &$   279.8\pm     2.4\pm     2.2$ &$   273.8\pm5.7\pm3.0     $\\
 3.8557  &$   318.8\pm     2.5\pm     2.6$ &$   317.8\pm6.2\pm3.5     $\\
 3.8636  &$   302.3\pm     2.5\pm     2.4$ &$   300.6\pm6.0\pm3.3     $\\
 3.8715  &$   514.2\pm     3.2\pm     4.1$ &$   507.7\pm7.8\pm5.6     $\\
 3.8805  &$   190.1\pm     2.0\pm     1.5$ &$   188.1\pm4.8\pm2.1     $\\
 3.8905  &$   184.1\pm     1.9\pm     1.5$ &$   172.2\pm4.6\pm1.9     $\\
\bottomrule
\end{tabular*}
\end{center}

To estimate the systematic uncertainties of the EMC cluster reconstruction and $E_{\rm{EMC}}$ requirement, we select a Bhabha sample with almost the same selection requirements as those listed in Section \textbf{4.1} except for the deposited energy requirement. Additional requirements of $E_{\rm{EMC}}>1.0~\rm{GeV}$ and $E_{\rm{EMC}}/p>0.8$ are imposed on one charged track and the other charged track is kept as the control sample. The difference of the acceptance efficiencies of the EMC cluster reconstruction and $E_{\rm{EMC}}$ requirement between data and MC simulation are taken as the systematic uncertainties.

The uncertainty of the trigger efficiency is less than 0.1\%~\cite{eff_trig}. The systematic uncertainty due to background is negligible.
The uncertainty associated with the signal MC model
due to the Babayaga generator is assigned to be $0.5\%$ according to Ref.~\cite{Babayaga_syserr}. To estimate the systematic uncertainty due to beam energy, we repeat the measurement by shifting the CM energies by $\pm 0.5$, $\pm 1$
 or $\pm 2\rm~MeV$, individually. The largest change in total integrated luminosity with respect to the nominal value is assigned as the
systematic uncertainty.

All of the systematic uncertainties are summarized in Table~\ref{table:syserr}.
Assuming the individual uncertainties to be independent, the total systematic uncertainty, 0.8\%, is calculated by adding them in quadrature.
\begin{flushleft}
\tabcaption{ \label{table:syserr}  Summary of systematic uncertainties in the luminosity measurement using the processes $e^+e^-\rightarrow(\gamma)e^+e^-$ and $e^+e^-\rightarrow(\gamma)\gamma\gamma$.}
\footnotesize
\begin{tabular*}{85mm}{c@{\extracolsep{\fill}}cc}
\toprule
\multirow{2}{*}{Source} &   \multicolumn{2}{c}{Systematic uncertainty (\%)}\\
& $e^+e^-\rightarrow(\gamma)e^+e^-$ & $e^+e^-\rightarrow(\gamma)\gamma\gamma$\\
\hline
$|\cos\theta|<0.70$                    &       0.2      & 0.2\\
Tracking and $p$ requirement           &       0.5      &  -   \\
$E_{\rm EMC}$ requirement              &       0.2      & 0.2  \\
EMC cluster reconstruction             &       0.06     & 0.06 \\
$\Delta\phi$ requirement               &       -        & 0.05 \\
Trigger efficiency                     &       0.1      & 0.1 \\
Generator                              &       0.5      & 1.0 \\
Beam energy                            &       0.11     & 0.11 \\
\hline
Total                                  &       0.8      & 1.1 \\
\bottomrule
\end{tabular*}
\end{flushleft}
\section{Cross check}
As a cross check, we perform an independent measurement of the integrated luminosities of the $\psi(3770)$ cross-section scan data by analyzing the process $e^+e^-\rightarrow(\gamma)\gamma\gamma$.

To select events from the process $e^+e^-\rightarrow(\gamma)\gamma\gamma$, we require that the number of good charged tracks is zero.
Two neutral clusters are required to be within the polar angle region $|\cos\theta|<0.7$ and the deposited energy
of each cluster in the EMC should be larger than $0.4\times E_{\rm b}$. Since the
directions of photons are not affected by the magnetic field, the two photon candidates should be back-to-back, and are required to satisfy $|\Delta\phi|<2.5^{\circ}$, where $\Delta\phi$ is defined as previously.
Figure~\ref{fig:ggdphi} shows a comparison of the $\Delta\phi$ distribution of
the $e^+e^-\rightarrow(\gamma)\gamma\gamma$ candidate events between the data taken at $E_{\rm cm}=3.7358$~GeV and the corresponding MC simulation. Good agreement is visible.
\begin{center}
\includegraphics[width=0.4\textwidth]{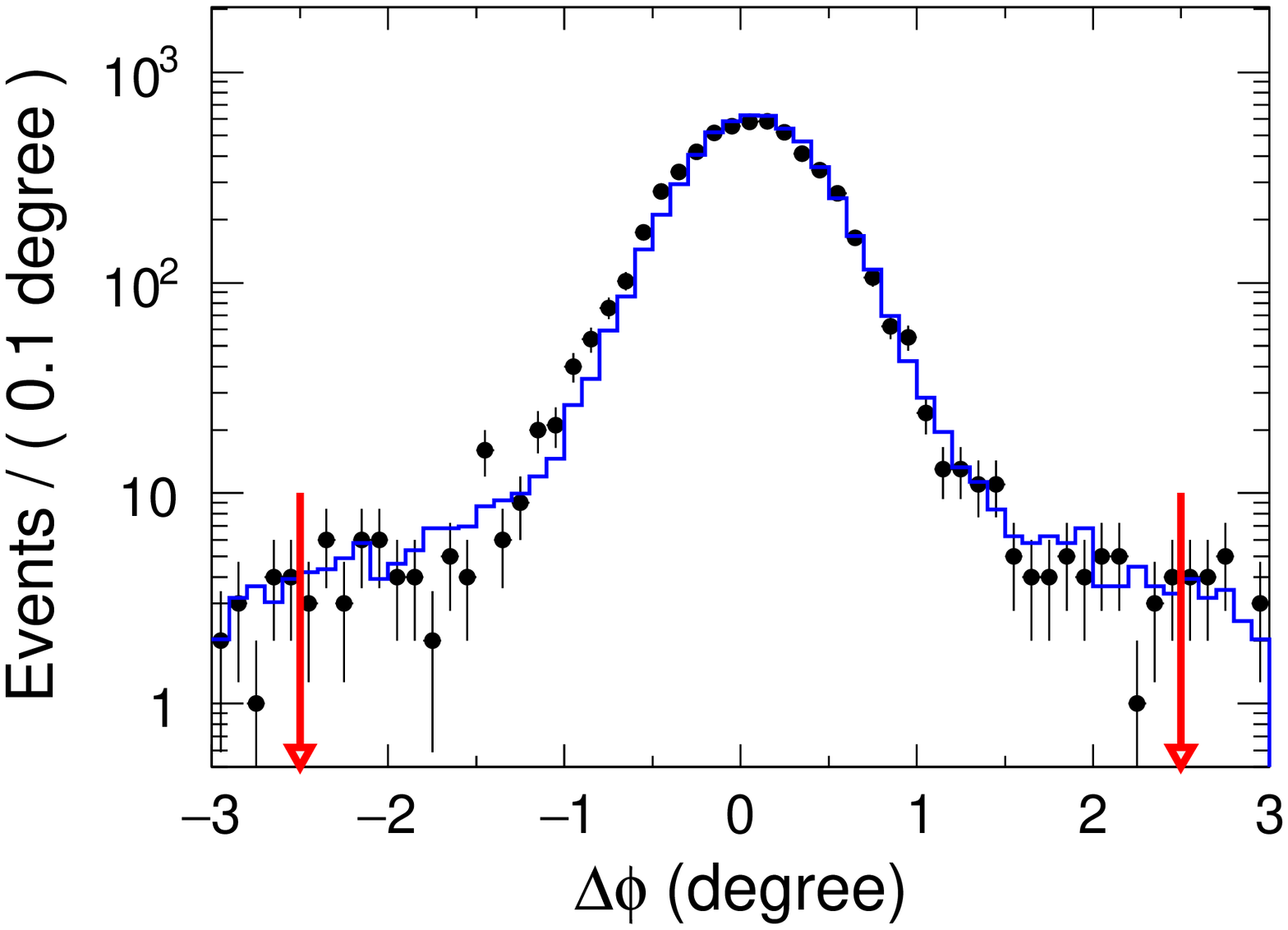}
\figcaption{\label{fig:ggdphi}  The $\Delta\phi$ distributions of the $e^+e^-\rightarrow(\gamma)\gamma\gamma$ candidate events selected from the data taken at $E_{\rm cm}=3.7358$~GeV (points with error bars) and the corresponding MC simulation (histogram). The selected $\Delta\phi$ range is indicated by the two arrows. The MC entries are normalized to the experimental data.}
\end{center}

For the background estimation, we analyzed the separated-beam data samples collected at 3.400 GeV and 4.030 GeV with BESIII, as well as MC samples of $\psi(3770)\to D\bar{D}$, $\psi(3770)\to$ non-$D\bar{D}$, $e^+e^-\to (\gamma)J/\psi$, $(\gamma)\psi(3686)$, $q\bar{q}$, $(\gamma)e^+e^-$, $(\gamma)\mu^+\mu^-$, and $(\gamma)\tau^+\tau^-$. The total contamination rate is estimated to be at the level of $10^{-3}$.

The integrated luminosity for the individual CM energy points is determined with Eq.~(\ref{eq:lum}) by using the numbers of observed $e^+e^-\to(\gamma)\gamma\gamma$ events, the contamination rates of backgrounds, the corresponding detection efficiencies, and cross sections calculated with the Babayaga v3.5 generator~\cite{babayaga}, as summarized in the third column of Table~\ref{tab:rslt}.
The main sources of the systematic uncertainty arise from the EMC cluster reconstruction, the requirements on $|\cos\theta|$, $E_{\rm EMC}$ and $\Delta\phi$, the trigger efficiency, the generator, and the beam energy. Most sources are the same as those in the luminosity measurement using Bhabha scattering events, and the corresponding systematic uncertainties are determined with the same approach.
To estimate the systematic uncertainty originating from the requirement on $\Delta\phi$, which is only used in the selection of $e^+e^-\rightarrow(\gamma)\gamma\gamma$ events, we repeat the measurements with the alternative requirements $|\Delta\phi|<2^{\circ}$ or $|\Delta\phi|<3^{\circ}$, individually. The maximum relative change of the integrated luminosity with respect to the nominal value is taken as the systematic uncertainty. The individual uncertainties are summarized in Table~\ref{table:syserr}, and the total systematic uncertainty, 1.1\%, is obtained by assuming the different systematic sources independently and adding the individual values in quadrature. The total integrated luminosity measured using $e^+e^-\rightarrow(\gamma)\gamma\gamma$ events is $75.50\pm0.09\pm0.83$~pb$^{-1}$, which is consistent with the result obtained using $e^+e^-\rightarrow(\gamma)e^+e^-$ within uncertainties, but with relatively larger statistical and systematical uncertainties.
\section{Summary}
\label{sec:sum}
By analyzing $e^+e^-\to(\gamma)e^+e^-$ events,
we measure the integrated luminosities of the $\psi(3770)$ cross-section scan data taken at 41 CM energy points. The total integrated luminosity of the $\psi(3770)$ cross-section scan data is determined to be $76.16\pm0.04\pm0.61$~pb$^{-1}$, where the first uncertainty is statistical and the second systematic. As a cross check, we also perform a measurement of the integrated luminosity for the $\psi(3770)$ cross-section scan data using $e^+e^-\to(\gamma)\gamma\gamma
$ events.  The results are consistent with that of the previous measurement, but with relatively larger uncertainty. The obtained integrated luminosities at the individual CM energy points are summarized in Table~\ref{tab:rslt}. The results provide important information needed to measure the cross sections of exclusive or inclusive hadronic production in
 $e^+e^-$ annihilation and thus benefit the understanding of the anomalous line-shape of $e^+e^-\to$ inclusive hadrons observed at BESII, the nature of the $\psi(3770)$, and the origin of the large branching fraction of $\psi(3770)\to$ non-$D\bar D$ decays~\cite{Int_ref2}.
\section{Acknowledgement}
The BESIII collaboration thanks the staff of BEPCII and the computing center for their hard efforts. This work is supported in part by National Key Basic Research Program of China under Contract No. 2015CB856700; National Natural Science Foundation of China (NSFC) under Contracts Nos. 11235011, 11335008, 11425524, 11625523, 11635010; the Chinese Academy of Sciences (CAS) Large-Scale Scientific Facility Program; the CAS Center for Excellence in Particle Physics (CCEPP); Joint Large-Scale Scientific Facility Funds of the NSFC and CAS under Contracts Nos. U1332201, U1532257, U1532258; CAS Key Research Program of Frontier Sciences under Contracts Nos. QYZDJ-SSW-SLH003, QYZDJ-SSW-SLH040; 100 Talents Program of CAS; National 1000 Talents Program of China; INPAC and Shanghai Key Laboratory for Particle Physics and Cosmology; German Research Foundation DFG under Contracts Nos. Collaborative Research Center CRC 1044, FOR 2359; Istituto Nazionale di Fisica Nucleare, Italy; Koninklijke Nederlandse Akademie van Wetenschappen (KNAW) under Contract No. 530-4CDP03; Ministry of Development of Turkey under Contract No. DPT2006K-120470; National Science and Technology fund; The Swedish Research Council; U. S. Department of Energy under Contracts Nos. DE-FG02-05ER41374, DE-SC-0010118, DE-SC-0010504, DE-SC-0012069; University of Groningen (RuG) and the Helmholtzzentrum fuer Schwerionenforschung GmbH (GSI), Darmstadt; WCU Program of National Research Foundation of Korea under Contract No. R32-2008-000-10155-0.
\end{multicols}

\vspace{-1mm}
\centerline{\rule{80mm}{0.1pt}}
\vspace{2mm}

\begin{multicols}{2}

\end{multicols}

\clearpage

\end{CJK*}
\end{document}